\documentstyle[aps,twocolumn]{revtex}

\begin{document}

\title{The beryllium atom and beryllium positive ion in strong
magnetic fields}
\author{M. V. Ivanov\dag\  and P. Schmelcher}
\address{Theoretische Chemie, Physikalisch--Chemisches Institut,
Universit\"at Heidelberg, INF 229, D-69120 Heidelberg,
Federal Republic of Germany\\
\dag Permanent address: Institute of Precambrian Geology and Geochronology,
Russian Academy of Sciences,
Nab. Makarova 2, St. Petersburg 199034, Russia
}

\date{\today}
\maketitle

\begin{abstract}
The ground and a few excited states of the beryllium atom
in external uniform magnetic fields are calculated
by means of our 2D mesh Hartree-Fock method for field strengths
ranging from zero up to $2.35\cdot 10^9$T.
With changing field strength the ground state
of the Be atom undergoes three
transitions involving four different electronic configurations
which belong to three groups with different spin projections
$S_z=0,-1,-2$. For weak fields the ground state configuration arises from the
$1s^2 2s^2 $, $S_z=0$ configuration.
With increasing field strength the ground state evolves into the
two $S_z=-1$ configurations
$1s^22s 2p_{-1}$ and
$1s^2 2p_{-1}3d_{-2}$,
followed by the fully spin polarised $S_z=-2$ configuration
$1s2p_{-1}3d_{-2}4f_{-3}$.
The latter configuration forms the ground state of the beryllium atom
in the high field regime $\gamma>4.567$.
The analogous calculations for the ${\rm Be^+}$ ion
provide the sequence
of the three following ground state configurations:
$1s^22s$ and
$1s^22p_{-1}$ ($S_z=-1/2$)
and $1s2p_{-1}3d_{-2}$ ($S_z=-3/2$).
\end{abstract}

\section{Introduction}

The behaviour and properties of atoms in strong magnetic fields
is a subject of increasing interest.
This is 
motivated by the astrophysical discovery of strong fields
on white dwarfs and neutron stars \cite{NStar1,NStar2,Whdwarf1}.
On the other hand the competition of the diamagnetic and Coulomb
interaction causes a rich variety of complex properties which are,
of course, also of interest on their own.

For a long time the investigations in the literature
focused on the hydrogen atom
(for a list of references see, for example,
\cite{RWHR,Ivanov88,Fri89,Kra96}).
As a result of the corresponding investigations
the absorption features of certain magnetic white dwarfs
could be understood in detail and a modelling
of their atmospheres was possible (see ref.\cite{Rud94} for a
review up to 1994 and \cite{Schm98} for more recent references).
Detailed spectroscopic calculations were carried out recently
for the helium atom in strong magnetic fields \cite{Bec99}.
These calculations allow to identify
spectra of other, namely helium-rich objects, including
the prominent white dwarf GD229 \cite{JSBS229}.
Recently a number of new magnetic white dwarfs
have been found
whose spectra are still unexplained
(see, e.g., Reimers et al \cite{Reim98}
in the course of the Hamburg ESO survey).

Investigations on the electronic structure in the presence of
a magnetic field appear to be quite complicated
due to the intricate geometry of this quantum problem.
For the hydrogen atom the impact of the competing Coulomb and diamagnetic interaction
is particularly evident and pronounced in the intermediate regime for which the magnetic
and Coulomb forces are comparable.
For different electronic degrees of excitation of the atom the intermediate
regime is met for different absolute values of the field strength.
For the ground state this regime corresponds to field strengths around $\gamma=~1$
(for the magnetic field strength as well as for other
physical values we use atomic units and, in particular,
$\gamma =B/B_0$, $B_0$
corresponds to the magnetic field strength $B_0=\hbar c/ea_0^2=2.3505 {\cdot} 10^5$T).
Both early \cite{Garstang,SimVir78} and more recent works
\cite{RWHR,Friedrich}
on the hydrogen atom have used different approaches for relatively
weak fields (the Coulomb force prevails over the magnetic force)
and for very strong fields (the Coulomb force can be
considered as weak in comparison with the magnetic forces which is the so-called
adiabatic regime).
A powerful method to obtain comprehensive results
on low-lying energy levels of the
hydrogen atom in particular in the intermediate regime is provided by
mesh methods \cite{Ivanov88}.
For atoms with several electrons there are two decisive factors which
enrich the possible changes in the electronic structure with varying
field strength compared to the one-electron system.
First we have a third competing interaction which is
the electron-electron repulsion and second the different electrons
feel very different Coulomb forces,
i.e. possess different one particle energies,
and consequently the regime of the intermediate field strengths
appears to be the sum of the intermediate regimes for the separate electrons.

Opposite to the hydrogen atom the wavefunctions of the multi-electron atoms
change their symmetries with increasing field strength.
It is well known that the singlet zero-field ground state
of the helium atom ($1s^2$ in the Hartree-Fock language) is replaced in
the high-field regime by the triplet
fully spin polarised configuration $1s2p_{-1}$.
For atoms with more than two electrons the evolution of the ground state
within the whole range of field strengths $0 \leq \gamma <+\infty$
includes multiple intermediate
configurations besides the zero-field ground state and
the ground state corresponding to the high field limit.
In view of the above there is a need for further quantum mechanical
investigations and data on atoms with more than two electrons
in order to understand their electronic structure
in strong magnetic fields.
Our calculations allowed
us to obtain the first conclusive results
on the series of ground state configurations
for the Li \cite{IvaSchm98} and C \cite{IvaSchm99} atoms.
These results are substantially different from previously
published ones \cite{JonesOrtiz}.
The ground state electronic configurations of the beryllium atom
for $0 \leq \gamma <+\infty$ were not investigated so far.
A previous work on the beryllium atom \cite{Ivanov98}
focused on problems
associated with the symmetries of the Hartree-Fock wavefunction
of the low-field ground state $1s^22s^2$ of this atom.
For strong fields the $1s^22s^2$ state represents a highly
excited state and the electronic ground state configuration
of Be is, so far, not investigated.

In the current paper we present results of our
fully numerical 2D Hartree-Fock mesh calculations
of the beryllium atom and ${\rm Be^+}$ ion
in magnetic fields and obtain for the first
time conclusive results on the structure and energy
of the ground state configurations of these systems for arbitrary
field strengths.

\section{Method}

The computational method applied in the current work coincides
with the method described in our works
\cite{Ivanov88,Ivanov98,Ivanov91,Ivanov94,ZhVychMat} and applied
afterwards in \cite{IvaSchm98,IvaSchm99,Schm98b,IvaSchm2000}.
We solve the electronic Schr\"odinger equation for the beryllium atom in
a magnetic field under the assumption of an infinitely heavy nucleus
in the (unrestricted) Hartree-Fock (HF) approximation.
The solution is established in the cylindrical coordinate system
$(\rho,\phi,z)$ with the $z$-axis oriented along the magnetic field.
We prescribe to each electron a definite value of the magnetic
quantum number $m_\mu$.
Each one-electron wave function $\Psi_\mu$ depends on the variables
$\phi$ and $(\rho,z)$
\begin{eqnarray}
\Psi_\mu(\rho,\phi,z)=(2\pi)^{-1/2}e^{-i m_\mu\phi}\psi_\mu(z,\rho)
\label{eq:phiout}
\end{eqnarray}
where $\mu$ indicates the numbering of the electrons.
The resulting partial differential equations for $\psi_\mu(z,\rho)$
and the formulae for the Coulomb and exchange potentials
have been presented in
ref.\cite{Ivanov94}.
These equations as well as the Poisson equations for inter-electronic
Coulomb and exchange potentials
are solved
by means of the fully numerical mesh method described in refs.
\cite{Ivanov88,Ivanov94}.
The finite-difference solution of the Poisson equations
on sets of nodes coinciding with those of the Hartree-Fock
equations turns out to be possible due to a special form
of uniform meshes used in the present calculations and
in refs.\cite{IvaSchm98,IvaSchm99,Ivanov98}.
Details and discussion on these meshes are presented
in ref.\cite{IvaSchmAdv}.

Our mesh approach is flexible enough to yield precise
results for arbitrary field strengths.
Some minor decrease of the precision
appears for electronic configurations with
big differences in the spatial distribution
of the electronic density for different electrons.
This results in big differences with respect
to the spatial extension of the density
distribution for different electrons.
This situation is more typical for the electronic configurations
which do not represent the ground state at
the corresponding fields
(e.g. $1s^22s^2$ at very strong fields or
$1s2p_{-1}3d_{-2}4f_{-3}$ in the weak field regime).
The precision of our results depends, of course, on the number of mesh nodes
and can be always improved in calculations with denser meshes.
Most of the present calculations
are carried out on sequences of meshes with the maximal number of nodes being
$80 \times 80$.

Along with the numerical solution of the Schr\"odinger equation
the key element for solving the problem of the ground state electronic configurations
is a proper choice of the configurations, which could
potentially be the ground state ones.
An example of solving this problem is presented in \cite{IvaSchm99}.
In that work we have developed a strategy which enables one to reduce
the set
of possible ground state configurations which are then subject
to a following numerical investigation.
This removes the risk of missing some ground state configurations
due to the limited possibilities of performing numerical investigations.
With increasing number of electrons the number of configurations
which cannot a priori be excluded from becoming the ground state
increase rapidly.
A comprehensive numerical investigation of all these configurations
is, in general, not feasible.
The above-mentioned strategy to exclude certain configurations
is therefore highly desirable.
It is based on a combination of qualitative theoretical
arguments and numerical calculations of the energies
of electronic configurations.
As a first step the set of electronic configurations has to be separated
into several groups according to their spin projections $S_z$.
The following considerations have to be carried out in each subset separately,
and are certainly more transparent by starting with the limit
of infinite strong fields and analysing the electronic configurations
with decreasing field strengths.
The qualitative theoretical considerations mentioned above
are based on the geometry of the spatial part
of the wavefunction and enable one to determine
the ground state for the high-field limit as well as
several candidates for the ground state
configuration with decreasing field strength.
The numerical calculations then enable us to decide which of
these candidates becomes the actual first intermediate ground state and
yields the transition field strength.
The knowledge of the first intermediate ground state allows us to repeat
the qualitative considerations for the second intermediate ground state
to obtain a list of candidates which is then investigated
by means of numerical calculations.
Repeating this procedure one can determine the full sequence of the
ground state configurations for each subset $S_z$ and
finally the sequence of ground state configurations 
for the physical system.

\section{Ground state electronic configurations for
$\gamma=0$ and $\gamma\rightarrow\infty$}

In this section we provide some helpful qualitative
considerations on the problem of the atomic multi-electron ground states
particularly in the limit of strong magnetic fields.

For the case $\gamma=0$ the ground state configuration 
of the beryllium atom can be characterised as $1s^2 2s^2$. 
This notation has a literal meaning when considering the atom 
in the framework of the restricted Hartree-Fock approach. 
The latter is an approximation of limited quality 
in describing the beryllium atom 
as it was shown in 
many fully correlated calculations
both for the field-free Be atom
\cite{Miller,SimsHagstrom} and for its polarizabilities in electric
fields \cite{DierSadj,Maroulis}.
It was pointed out in these works 
that the Be atom is a strongly correlated system and
that the HF ground state wavefunction 
(i. e. the spherically symmetric $1s^2 2s^2$) 
is not a very accurate zeroth-order wavefunction, especially for calculations
of electric polarizabilities. 
This is due to a significant contribution of the
$1s^22p^2$ configuration to the ground state wave function. 
The latter configuration is evidently a non-spherical one. 
This fact is in agreement with results of ref. \cite{Ivanov98}
where the fully numerical 2D unrestricted Hartree-Fock approach provides 
the $2s^2$ shell stretched along the $z$ axis even for $\gamma=0$. 
In terms of spherical functions it is natural to describe 
this geometry of the $2s^2$ shell 
as a mixture of $2s$ and $2p_0$ functions. 
We remark that the $s$, $p$, $d\ldots$ orbital notation 
both for $\gamma=0$ and $\gamma\neq 0$ 
is based on the behaviour of the wave functions 
in the vicinity of the origin 
and on the topology of the nodal surfaces,
but does not imply any detailed geometry 
or certain values of the orbital moment $l$.

It is evident that the field-free ground state 
of the beryllium atom remains the
ground state only for relatively weak fields.
The set of one-electron wave functions constituting
the HF ground state for the opposite case of extremely strong
magnetic fields can be determined as follows.
The nuclear attraction energies and HF potentials
(which determine the motion along $z$ axis)
are small for large $\gamma$ in comparison
to the interaction energies with the magnetic field
(which determines the motion perpendicular to the magnetic field
and is responsible for the Landau zonal structure of the spectrum).
Thus, all the one-electron wavefunctions must correspond to the
lowest Landau zones, i.e. 
the magnetic quantum numbers $m_\mu$ are not postive for 
all the electrons 
$m_\mu\leq 0$,
and the system must be fully spin-polarised,
i.e. $s_{z\mu}= -{1\over2}$.
For the Coulomb central field the one-electron levels form
(as $B\rightarrow\infty$) quasi 1D Coulomb series with the binding energy
$\epsilon_B={1\over{2n_z^2}}$ for $n_z>0$ and
$\epsilon_B\rightarrow \infty$ for $n_z=0$,
where $n_z$ is the number of nodal
surfaces of the wave function with respect to the $z$ axis.
The binding energy of a separate electron has the form 
\begin{eqnarray}
\epsilon_{{\rm B}}=(m+|m|+2s_{z}+1)\gamma/2-\epsilon
\label{eq:ebin1}
\end{eqnarray}
where $\epsilon$ is the energy of the electron.

When considering the case $\gamma \rightarrow \infty$ 
it is evident, that the wave functions
with $n_z=0$ have to be chosen for the ground state configuration.
Furthermore starting with the energetically lowest one particle
level the electrons occupy according to the above arguments 
orbitals with increasing
absolute value of the magnetic quantum number $m_{\mu}$.
Consequently the ground state of the beryllium atom
must be given by the fully spin-polarised configuration
$1s2p_{-1}3d_{-2}4f_{-3}$.
In our notation of the electronic
configurations we assume in the following
that all paired electrons, like
for example the $1s^2$ part of a configuration, are of course in a
spin up and spin down orbital, respectively, whereas all
unpaired electrons possess a negative projection of the
spin onto the magnetic field direction.
On a qualitative level the configuration $1s2p_{-1}3d_{-2}4f_{-3}$ 
is not very different from similar electronic configurations 
for other atoms (see ref. \cite{IvaSchm2000}). 
This is a manifestation of the simplification of the picture 
of atomic properties in the limit $\gamma\rightarrow\infty$ 
where a linear sequence of electronic configurations replaces 
the periodic table of elements of the field-free case.

The problem of the configurations of the
ground state for the intermediate field region cannot be
solved without doing explicit calculations
combined with some qualitative considerations in order to
extract the relevant configurations.

\section{Ground state electronic configurations
for arbitrary field strengths}

In order to determine the ground state electronic configurations
of the beryllium atom
we employ here the strategy introduced in ref.\cite{IvaSchm99}
where the carbon atom has been investigated.
First of all, we divide
the possible ground state configurations into three groups
according to their total spin projection $S_z$ :
the $S_z=0$ group (low-field ground state configurations),
the intermediate group $S_z=-1$ and the
$S_z=-2$ group (the high-field ground state configurations).
This grouping is required for the following qualitative considerations 
which are based on the geometry
of the spatial parts of the one-electron wave functions.
In the course of this discussion
it is expedient to treat
{\it local} ground states for each $S_z$ subset
(i.e. the lowest states with a certain $S_z$ value)
along with the {\it global} ground state
of the atom as a whole.
For each value of the magnetic field strength one of these
local ground states is the global ground state of the atom.

According to the general arguments presented in the previous section
we know that the ground state configuration of the beryllium atom
in the high field limit must be the fully spin-polarised state
$1s2p_{-1}3d_{-2}4f_{-3}$.
The question of the ground state configurations at intermediate fields
cannot be solved without performing explicit electronic structure calculations.
On the other hand, the a priori set of possible intermediate ground
state configurations increases enormously with increasing number of electrons
and is rather large already for the beryllium atom.
Some qualitative considerations are therefore
needed in order to exclude certain configurations as possible ground
state configurations thereby reducing the number of candidates for which
explicit calculations have to be performed.
As mentioned in the previous section
the optimal strategy hereby consists of the repeated procedure
of determining neighbouring ground state configurations with
increasing (or decreasing) magnetic field strength using both qualitative
arguments as well as the results
of the calculations for concrete configurations.

The total energies for the considered states and
particularly of those states which become the global
ground state of the atom for some regime of the field
strength are illustrated in figure 1.
In the following paragraphs we describe
our sequence of selection procedure and calculations for the candidates
of the electronic ground state configurations.

Due to the simplicity of the ground state electronic configurations
of atoms in the limit $\gamma\rightarrow\infty$
it is natural to start the consideration for $\gamma\ne 0$  
with the high-field ground state and
then consider other possible candidates in question
for the electronic ground state for $S_z=-2$
(see figure 1) {\it{with decreasing field strength}}. 
The consideration of the high-field 
(i.e. the fully spin-polarised) regime was carried out 
in ref. \cite{IvaSchm2000} and for this case 
(i.e. $S_z=-2$ for beryllium) 
we repeat this consideration in more detail. 
In particular, we have found in ref. \cite{IvaSchm2000} 
that the beryllium atom, opposite to the carbon and heavier elements 
has only one fully spin-polarised ground state configuration.

All the one electron wave functions of the high-field ground state
$1s2p_{-1}3d_{-2}4f_{-3}$
possess no nodal surfaces crossing the $z$-axis and 
occupy the energetically lowest orbitals
with magnetic quantum numbers ranging from $m=0$ down to $m=-3$.
The $4f_{-3}$ orbital possesses the smallest binding energy 
of all orbitals constituting
the high-field ground state. 
Its binding energy decreases rapidly with decreasing
field strength.
Thus, we can expect that the first crossover 
of ground state configurations
happens due to a change of the $4f_{-3}$ orbital into one
possessing a higher binding energy at the corresponding
lowered field strength.
One may think that the first transition while 
decreasing the magnetic field strength
will involve a transition from an orbital 
possessing $n_z=0$ to one for $n_z=1$.
The energetically lowest available one particle state
with $n_z=1$ is the $2p_0$ orbital.
Another possible orbital into which the $4f_{-3}$
wave function could evolve is the $2s$ state.
For the hydrogen atom or hydrogen-like ions
in a magnetic field the $2p_{0}$ is stronger bound than the
$2s$ orbital.
On the other hand, owing to the electron screening 
in multi-electron atoms in field-free space
the $2s$ orbital tends to be more tightly bound than the $2p_0$ orbital.
Thus, two states i.e. $1s2p_02p_{-1}3d_{-2}$ and
$1s2s2p_{-1}3d_{-2}$ are candidates for becoming the ground state
in the $S_z=-2$ set when we lower the field strength 
coming from the high field situation.
The numerical calculations show that
the first crossover of the
$S_z=-2$ subset takes place between the
$1s2p_{-1}3d_{-2}4f_{-3}$ and $1s2p_02p_{-1}3d_{-2}$
configurations (figure 1).
On the other hand, the calculations show that even earlier
(i.e. at higher magnetic field strengths)
the global ground state acquires the total spin $S_z=-1$
due to a crossover of the energy curve of the
$1s2p_{-1}3d_{-2}4f_{-3}$ configuration with that of the
configuration $1s^22p_{-1}3d_{-2}$ (which is the local ground state for
the $S_z=-1$ subset in the high-field limit).
For the fields below this point $\gamma=4.567$ the ground state
electronic configurations of the beryllium atom belong
to the subset $S_z=-1$.
This means that the beryllium atom has only one fully spin
polarised ground state configuration (as mentioned above).

The electronic configurations $1s^22p_{-1}3d_{-2}$ and
$1s2p_{-1}3d_{-2}4f_{-3}$ differ by the replacement
of the spin down $4f_{-3}$ orbital through the spin up
$1s$ orbital and according to the argumentation
presented in the previous section
the $1s^22p_{-1}3d_{-2}$ represents the {\it local}
ground state configuration for the subset $S_z=-1$ in the limit
$\gamma\rightarrow\infty$.
Analogous arguments to that presented
in the previous paragraph provide the conclusion, that
in the process of decreasing field strength
the $1s^22p_{-1}3d_{-2}$ ground state electronic 
configuration can be replaced
either by the $1s^22s2p_{-1}$ or by the $1s^22p_02p_{-1}$
configuration.
The numerical calculations show, 
that the curve $E_{1s^22s2p_{-1}}(\gamma)$
intersects the curve $E_{1s^22p_{-1}3d_{-2}}(\gamma)$
at a higher magnetic field ($\gamma=0.957$)
than $E_{1s^22p_02p_{-1}}(\gamma)$ crosses 
$E_{1s^22p_{-1}3d_{-2}}(\gamma)$.
The difference with respect to the order of the local ground state
configurations in the subsets $S_z=-2$ and $S_z=-1$ 
stems from the difference in the magnetic field strengths
characteristic for the crossovers in these subsets.
At moderate field strengths ($S_z=-1$)
the influence of the Coulomb fields of the nucleus
and electrons prevails over the influence
of the magnetic field and make the energy of the $2s$ orbital
lower than that of the $2p_0$ orbital.
On the other hand, at stronger fields characteristic
for the subset $S_z=-2$ the energies of these orbitals
are governed mostly by the magnetic field and, in result,
the energy of the $2p_0$ orbital becomes lower than
the energy of the $2s$ orbital.

From our simple qualitative considerations we can conclude,
that the configuration $1s^22s2p_{-1}$ is the {\it local}
ground state configuration of the subset $S_z=-1$
for the weak field case, i.e. for $\gamma\rightarrow 0$.
Indeed, when we construct such a configuration, the
first three electrons go to orbitals $1s$ and $2s$
forming the $1s^22s$ configuration with $S_z=-1/2$.
The fourth electron must then have the same spin as the $2s$ 
orbital electron to obtain the total spin value $S_z=-1$.
Thus, the lowest orbital which it can occupy is the $2p_{-1}$.
Therefore, there are two local ground state configurations
in the subset $S_z=-1$ and they both represent
the global ground state for some ranges of the magnetic
field strengths.

The necessary  considerations for the subset $S_z=0$ are quite simple.
At $\gamma=0$ and, evidently, for very weak fields
the ground state of the beryllium atom has the configuration
$1s^22s^2$.
We can expect, that when increasing the magnetic field strength,
the next lowest state with $S_z=0$ will be 
the $1s^22s2p_{-1}$ configuration with
opposite directions of the spins of the $2s$ and $2p_{-1}$ electrons.
But both contributions, the Zeeman spin term and 
the electronic exchange make
the energy of this state higher than the energy of the state
$1s^22s2p_{-1}$ with the parallel orientation of the spins
of the $2s$ and $2p_{-1}$ electrons (i.e. $S_z=-1$) considered
above.
The calculated energies for these states are presented in
figure 1.
Thus, the beryllium atom has one ground state electronic
configuration $1s^22s^2$ with the total spin $z$-projection $S_z=0$.
This state is the global ground state for
the magnetic field strengths between $\gamma=0$ and $\gamma=0.0412$.
Above this point the ground state configuration is $1s^22s2p_{-1}$
with $S_z=-1$.

Summarising the results on the ground
state configurations of the beryllium atom we can state
that depending on the magnetic field strength this
atom has four different electronic ground state configurations.
For $0\leq\gamma<0.0412$ the ground state configuration
coincides with the field-free ground state
configuration $1s^22s^2$ which has zero values for the total
magnetic quantum number $M$ and spin projection $S_z$.
The following are two ground state configurations with $S_z=-1$:
$1s^22s2p_{-1}$ ($M=-1$) for $0.0412<\gamma<0.957$ and
$1s^22p_{-1}3d_{-2}$ ($M=-3$) for $0.957<\gamma<4.567$.
For $\gamma>4.567$ the ground state configuration is
$1s2p_{-1}3d_{-2}4f_{-3}$ with $S_z=-2$ and $M=-6$.
The complete results of the investigations of the sequence
of the ground state configurations of the Be atom
are presented in table \ref{tab:Begrcon} which contains
the critical values of $\gamma$
at which the crossovers of different
ground state configurations take place.

The next aim of this section is the corresponding
investigation of the ground state configurations of the
ion ${\rm Be^+}$.
The field-free ground state of this ion corresponds to the 
$1s^22s$ configuration ($S_z=-1/2$ and $M=0$).
In the opposite case $\gamma\rightarrow\infty$
the ground state is obviously given by the 
$1s2p_{-1}3d_{-2}$ configuration ($S_z=-3/2$ and $M=-3$).
Thus, we need to investigate only two different subsets of
electronic ground state configurations: $S_z=-1/2$ and $S_z=-3/2$.
The energy curves which are necessary for this investigation
are presented in figure 2.
The subset $S_z=-1/2$ contains only two possible
ground state configurations $1s^22s$ and $1s^22p_{-1}$.
The latter is the {\it local} ground state configuration
for this subset in the limit $\gamma\rightarrow\infty$.
The curves $E_{1s^22s}(\gamma)$ and $E_{1s^22p_{-1}}(\gamma)$
intersect at $\gamma=0.3185$ and above this point
$E_{1s^22p_{-1}}<E_{1s^22s}$.
In the subset $S_z=-3/2$ we have to consider the configurations
$1s2p_02p_{-1}$ and $1s2s2p_{-1}$ along with
the high-field ground state configuration $1s2p_{-1}3d_{-2}$.
But the numerical calculations show that the energies of
both $1s2p_02p_{-1}$ and $1s2s2p_{-1}$ lie above the
energy of the $1s2p_{-1}3d_{-2}$ configuration at
the intersection point ($\gamma=4.501$) between
$E_{1s2p_{-1}3d_{-2}}(\gamma)$ and $E_{1s^22p_{-1}}(\gamma)$.
Thus, the ion ${\rm Be^+}$ has three different electronic
ground state configurations in external magnetic fields:
for $0\leq\gamma<0.3185$ it is $1s^22s$ ($S_z=-1/2$ and $M=0$),
then for $0.38483<\gamma<4.501$ it is 
$1s^22p_{-1}$ ($S_z=-1/2$ and $M=-1$)
and for all the values $\gamma>4.501$ the ground state configuration is
$1s2p_{-1}3d_{-2}$ ($S_z=-3/2$ and $M=-3$).
These results are summarised in table \ref{tab:Bepgrcon}.
The set of the electronic ground state configurations
for the ${\rm Be^+}$ ion appears to be qualitatively the same
as for the lithium atom \cite{IvaSchm98}.
The field strengths for the corresponding transition points
are roughly two times higher for the ${\rm Be^+}$ ion
than for the Li atom.

\section{Selected quantitative aspects}

In tables \ref{tab:Beenergy} and \ref{tab:Bepenergy}
we present the total energies of the four
ground state electronic configurations of the beryllium atom
and the three ground state electronic configurations of the ion
${\rm Be^+}$, respectively.
These data cover a very broad range of the field strengths
from $\gamma=0$ and very weak magnetic fields starting with
$\gamma=0.001$ up to extremely strong fields $\gamma=10000$.
The latter value of the field strength can be considered as
a rough limit of applicability of the non-relativistic
quantum equations to the problem (see below).
The corresponding data on the ${\rm Be^+}$ ion can be found in
tables \ref{tab:Bepgrcon} and \ref{tab:Bepenergy}.

So far there exist three works which should be 
mentioned in the context of the problem
of the beryllium atom in strong magnetic fields.
Ref. \cite {Ivanov98} deals with the $1s^22s^2$
state of this atom in fields $0\leq\gamma\leq 1000$ and
ref. \cite{IvaSchm2000} investigates the ground state energies of
atoms with nuclear charge number $Z\leq 10$ in the high-field,
i.e. fully spin polarised regime.
Both these works contain calculations carried out by
the method used in the current work and do not represent
a basis for comparison.
The comparison of our results with an adiabatic Hartree-Fock calculation
of atoms with $Z\leq 10$ \cite{Neuhauser} is presented
in \cite{IvaSchm2000} and we can briefly summarise this
comparison for two values of the magnetic field strengths:
for $B_{12}=0.1$ (i.e. $B=0.1\times 10^{12}$G) our result is $E=-0.89833$keV 
whereas ref. \cite{Neuhauser} yields $E=-0.846$keV;
for $B_{12}=5$ (i.e. $B=5\times 10^{12}$G) our result is $E=-3.61033$keV,
whereas ref. \cite{Neuhauser} yields $E=-3.5840$keV.
This comparison allows us to draw the conclusion of
a relatively low precision of the adiabatic approximation for multi-electron
atoms even for relatively high magnetic fields.

In figure 3 we present the ionization energy
$E_{\rm ion}$ of the beryllium atom
depending on the magnetic field strength.
This continuous dependence is divided into six parts corresponding
to different pairs of the ground state configurations of the Be atom
and ${\rm Be^+}$ ion involved into the ionization energy.
The five vertical dotted lines in figure 3 mark the boundaries
of these sections.
The alteration of the sections of growing and decreasing
ionization energy originates from different dependencies
of the total energies of the Be and ${\rm Be^+}$ on the
magnetic field strength for different pairs
of the ground state configurations of these two systems.
One can see the sharp decrease of the ionization energy
between the crossovers (4) and (5).
This behaviour is due to the fact that $E_{\rm ion}$ is determined
in this section by the rapidly decreasing total energy of the state
$1s2p_{-1}3d_{-2}$ of the ${\rm Be^+}$ ion (figure 2)
and by the energy of the Be atom in the state $1s^22p_{-1}3d_{-2}$ 
which is very weakly dependent on the field strength (figure 1).
Another remarkable feature of the curve $E_{\rm ion}(\gamma)$
is its behaviour in the range of field strengths 
between the transitions (2) and (3).
The ionization energy in this region contains a very shallow
maximum and in the whole section it is almost independent
on the magnetic field.
Thus, 
the ionization energy is stationary in this regime of field strengths
$\gamma=0.3 - 0.5 $~a.u.
typical for many magnetic white dwarfs \cite{Rud94}.

The above-discussed properties are based on the behaviour
of the total energy of the Be atom and ${\rm Be^+}$ ion.
On the other hand, the behaviour of the wavefunctions
and many intrinsic characteristics
of atoms in external magnetic fields are associated not
with the total energy, but with the binding energies of
separate electrons (\ref{eq:ebin1}) 
and the total binding energy of the system
\begin{eqnarray}
E_{\rm B}=\sum_{\mu=1}^N(m_\mu+|m_\mu|+2s_{z\mu}+1)\gamma/2-E
\label{eq:ebintot}
\end{eqnarray}
where $N$ is the number of electrons. 
The binding energies of the ground state electronic configurations
of Be and ${\rm Be^+}$ depending
on the magnetic field strength are presented in figures 4 and 5.
These dependencies at very strong magnetic fields may illustrate
our considerations of the previous sections.
One can see in figure 4 that the high-field ground state
$1s2p_{-1}3d_{-2}4f_{-3}$ is not the most tightly bound state
of the beryllium atom.
For all the values of the magnetic fields considered in this paper
its binding energy is
lower than that of states $1s^22s2p_{-1}$ and $1s^22p_{-1}3d_{-2}$
and for $\gamma<100$ it is lower than $E_{{\rm B}1s^22s^2}$.
The latter circumstance can be easily explained by the fact
that the $1s^22s^2$ configuration contains two tightly
bound orbitals $1s$ whereas the $1s2p_{-1}3d_{-2}4f_{-3}$
possess only one such orbital.
However, with increasing magnetic field strengths the contribution
of the group $2p_{-1}3d_{-2}4f_{-3}$ to the binding energy
turns out to be larger than that of the $1s2s^2$ group.
Analogously we can expect
$E_{{\rm B}1s2p_{-1}3d_{-2}4f_{-3}}>E_{{\rm B}1s^22s2p_{-1}}$
at some very large fields $\gamma>10000$.
On the other hand, it is evident that the state
$1s2p_{-1}3d_{-2}4f_{-3}$ must be less bound than $1s^22p_{-1}3d_{-2}$
because both these configurations are constructed of
orbitals with binding energies, logarithmically increasing
as $\gamma\rightarrow\infty$, but the $1s^22p_{-1}3d_{-2}$
contains an additional $1s$ orbital, which is more tightly
bound than $4f_{-3}$ at arbitrary field strengths.
The plot for the ${\rm Be^+}$ ion (figure 5) illustrates
the same features and one can see in this figure nearly
parallel curves $E_{{\rm B}1s^22p_{-1}}(\gamma)$ and
$E_{{\rm B}1s2p_{-1}3d_{-2}}(\gamma)$ in the strong field regime.

Figures 6 and 7 allow us to add some details
to the considerations of the previous section.
These figures present spatial distributions of the total electronic densities
for the ground state configurations of the beryllium atom 
and its positive ion, respectively.
These pictures allow us to gain insights into the
geometry of the distribution of the electronic density 
in space and in particular
its dependence on the magnetic quantum number and the total spin.
The first pictures in these figures present the distribution of
the electronic density for the ground state
for $\gamma=0$.
The following pictures show the distributions of the electronic densities
for values of the field strength which mark 
the boundaries of the regimes of field
strengths belonging to the different ground state configurations.
For the high-field ground states we present the distribution of
the electronic density at the crossover field strength
and at $\gamma=500$.
For each configuration the effect of the increasing field strength
consists in compressing the electronic distribution towards the $z$ axis.
However the crossovers of ground state configurations
involve the opposite effect due to the fact that these crossovers
are associated with an increase of the total magnetic quantum number
$M$. 

In the first rows of figures 6 and 7 
one can see a dense core of $1s^2$ electrons 
inside the bold solid line contour 
and a diffuse distribution of $2s$ electrons outside this core. 
The prolate shape of the bold solid line contour in the first plot 
of the figure 6 ($1s^2 2s^2$, $\gamma=0$) 
reflects the non-spherical distribution of the $2s$ electrons 
in our calculations or the admixture of the $1s^2 2p_0^2$ configuration 
to the $1s^2 2s^2$ one from the point of view 
of the multi-configurational approach 
\cite{Miller,SimsHagstrom,DierSadj,Maroulis}.

Some additional issues concerning the results 
presented above have to be discussed.
First, our HF results do not include the effects of correlation. To take into account the latter
would require a multi-configurational approach which goes beyond
the scope of the present paper. We, however, do not expect that
the correlation energy changes our main conclusions like, for example, the
fact of the crossovers with respect to the different ground states configurations.
With increasing field strength the effective one particle
picture should be an increasingly better description of the wave function
and the percentage of the correlation energy should therefore
decrease (see ref.\cite{Schm98b} for an investigation on this subject).
Two other important issues are relativistic effects and
effects due to the finite nuclear mass.
Both these points are basically important for very
high magnetic field strengths
and they have been discussed in ref.\cite{IvaSchm2000}.
For the systems Be and ${\rm Be^+}$ 
and for most of the field strengths considered here 
these effects
result in minor corrections to the energy.

\section{Summary and conclusions}

We have applied our 2D mesh Hartree-Fock method to the magnetised
neutral beryllium atom and beryllium positive ion.
The method is flexible enough to yield precise results for arbitrary
field strengths and our calculations for the ground and several excited
states are performed for magnetic field strengths ranging from zero
up to  $2.3505\cdot 10^9$T ($\gamma=10000$).
Our considerations focused on the ground states and
their crossovers with increasing field strength.
The ground state of the beryllium atom
undergoes three transitions involving
four different electronic configurations.
For weak fields up to $\gamma=0.0412$ the ground state arises
from the field-free ground state configuration
$1s^2 2s^2$ with the total spin $z$-projection $S_z=0$.
With increasing strength of the field two
different electronic configurations with
$S_z=-1$ consequently become the ground state:
$1s^22s 2p_{-1}$ and $1s^22p_{-1}3d_{-2}$.
At $\gamma=4.567$ the last crossover of the ground state configurations
takes place and for $\gamma>4.567$ the ground state wavefunction is represented by the
high-field-limit fully spin polarised configuration
$1s2p_{-1}3d_{-2}4f_{-3}$, $S_z=-2$.

For the ion ${\rm Be^+}$ we obtain three different
ground state configurations possessing two values
of the spin projection.
For fields below $\gamma=0.3185$ the ground state electronic
configuration has the spin projection $S_z=-1/2$,
magnetic quantum number $M=0$ and qualitatively coincides with
the zero-field ground state configuration $1s^22s$.
Between $\gamma=0.3185$ and $\gamma=4.501$ the ground state
is represented by another configuration with $S_z=-1/2$, i.e.
$1s^22p_{-1}$ ($M=-1$).
Above the point $\gamma=4.501$ the fully spin
polarised high-field-limit configuration $1s2p_{-1}3d_{-2}$
($S_z=-3/2$) is the actual ground state of the ${\rm Be^+}$ ion.
Thus, the sequence of electronic ground state configurations
for the ${\rm Be^+}$ ion is similar to the sequence
for the Li atom \cite{IvaSchm98}.
We present detailed tables of energies of the ground state configurations 
for Be and ${\rm Be^+}$. 

For Be and ${\rm Be^+}$ we have presented also the binding 
energies of the ground state configurations dependent on the 
magnetic field strength and maps of electronic densities 
for these configurations.
For the Be atom we present its ionization energy dependent 
on the field strength. 

Our investigation represents the first conclusive study of the ground state
of the beryllium atom and ${\rm Be^+}$ ion for arbitrary field strengths.
For the Be atom we have obtained a new
sequence of electronic configurations with increasing field strength.
This sequence does not coincide with any such sequences obtained previously
for other atoms and ions and could not be predicted even qualitatively
without detailed calculations.
Putting together what we currently know about
ground states of atomic systems in strong magnetic fields
we can conclude that the H, He, Li, Be, C, ${\rm He^+}$, ${\rm Li^+}$ and ${\rm Be^+}$
ground states have been identified.
For other atoms and multiple series of ions
the question about the
ground state configurations is still open.

{}

\vspace*{2.0cm}

{\bf Figure Captions}

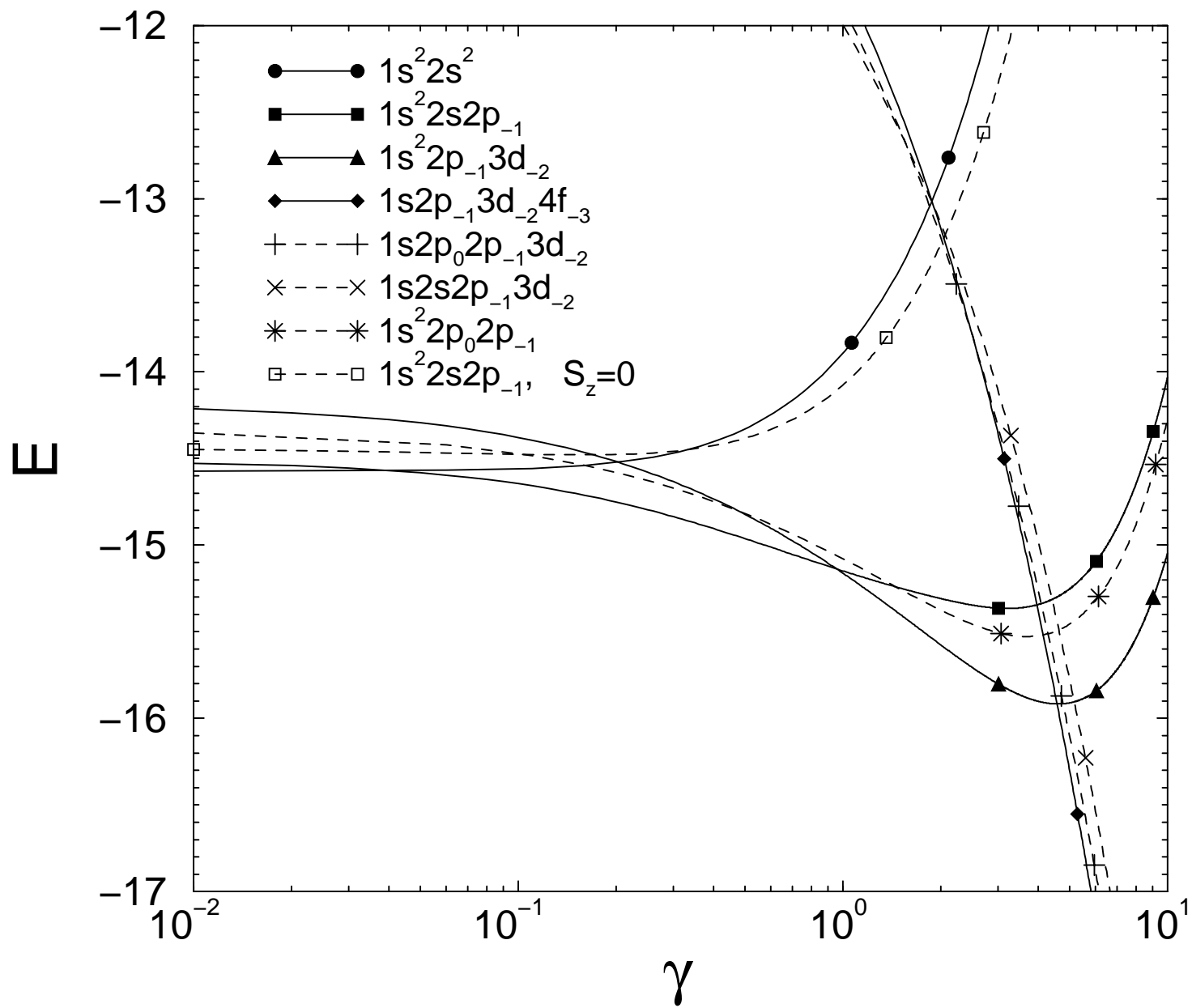
{\bf Figure 1.} The total energies (in atomic units) of the states
of the beryllium atom as functions
of the magnetic field strength
considered for the determination of the ground state electronic configurations.
The field strength is given in units of $\gamma=(\frac{B}{B_0})$,$B_0=\hbar c/ea_0^2=2.3505 {\cdot} 10^5$T.

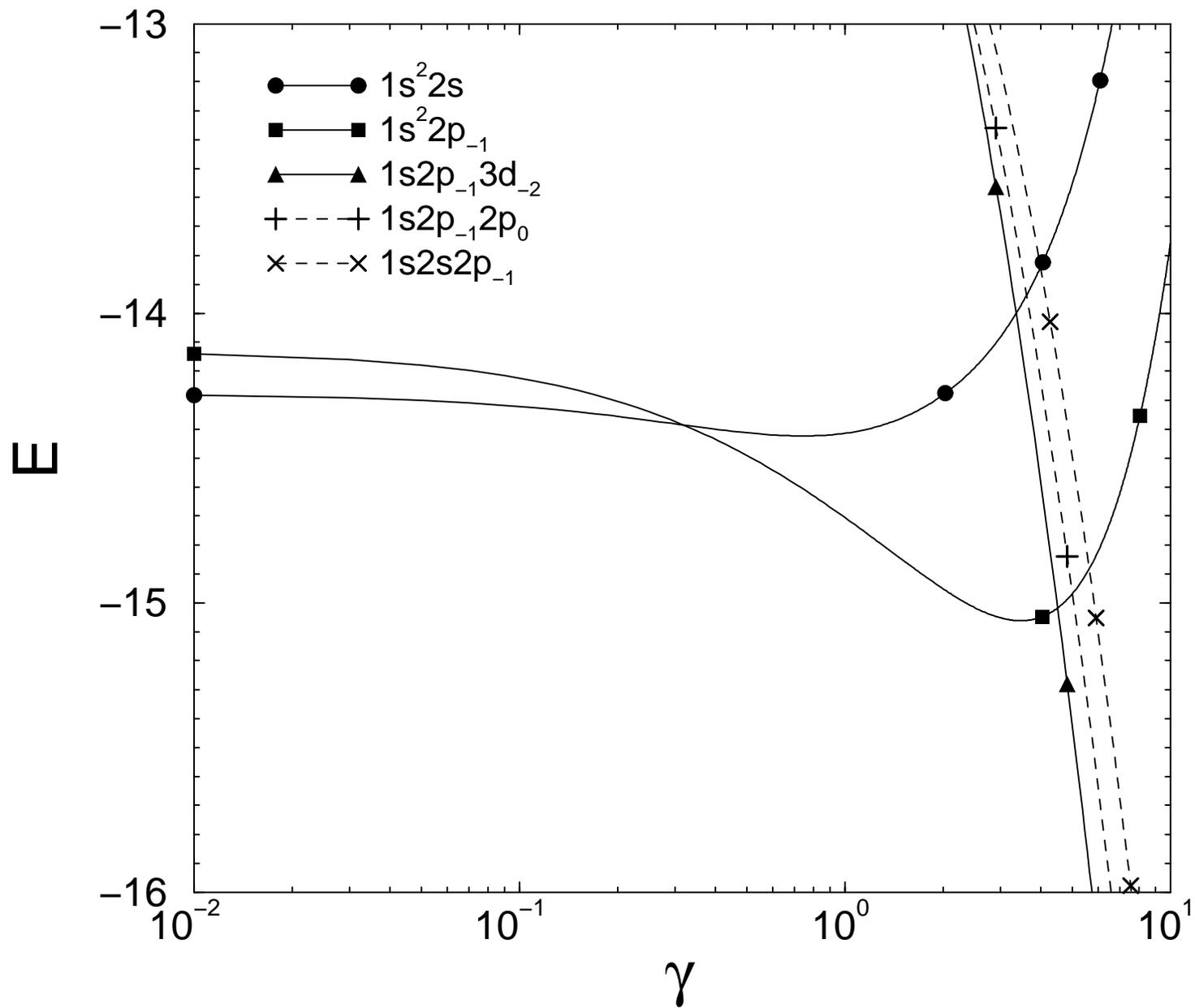
{\bf Figure 2.} The total energies (in atomic units) of the states
of the beryllium positive ion as functions
of the magnetic field strength
considered for the determination of the ground state electronic configurations.
The field strength is given in units of $\gamma=(\frac{B}{B_0})$,$B_0=\hbar c/ea_0^2=2.3505 {\cdot} 10^5$T.

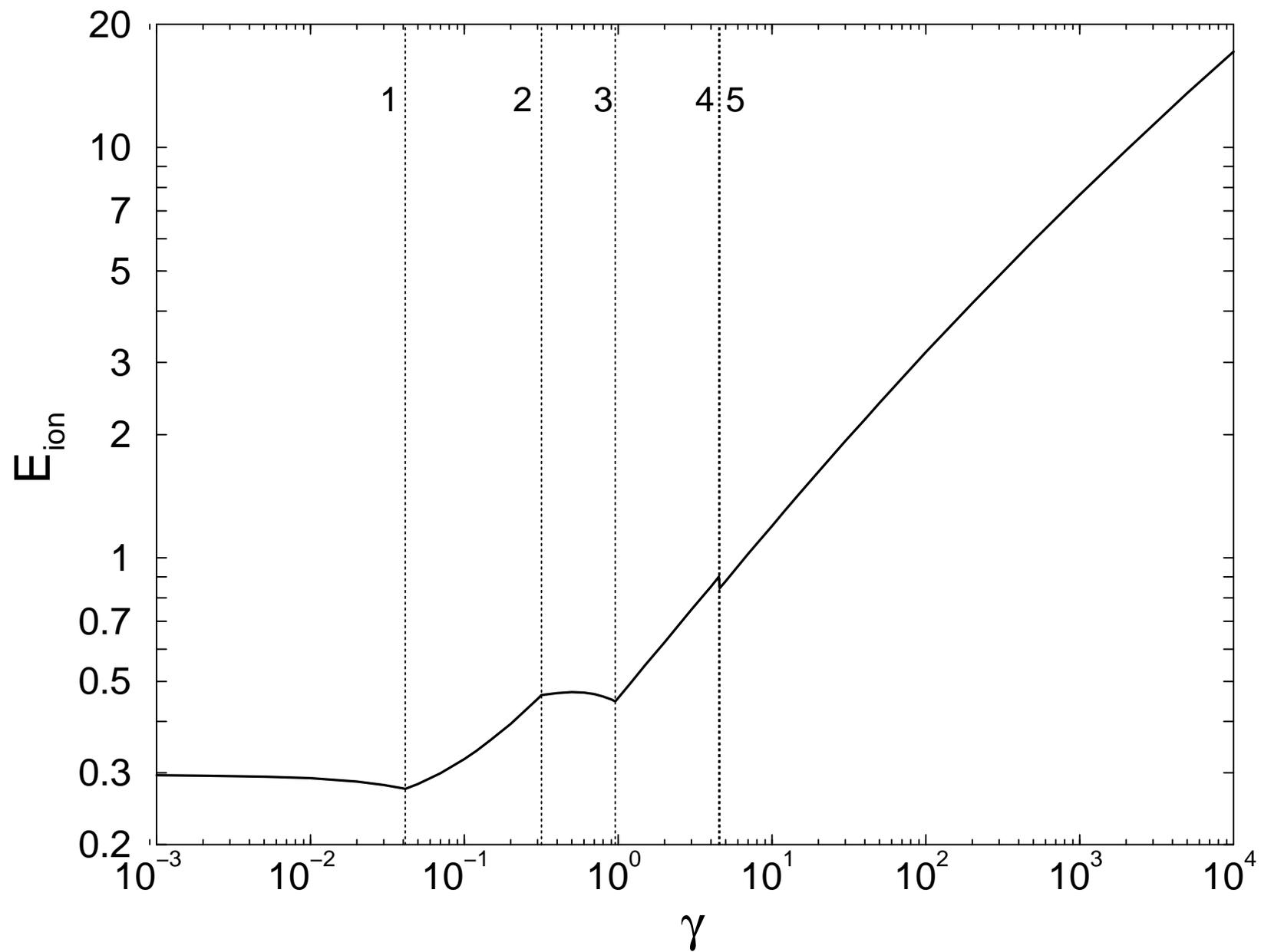
{\bf Figure 3.} Be atom ground state ionization energy $E_I$.
Transition points are marked by broken vertical lines.
The sequence of the transitions are (from left to right):
1. Be: $1s^22s^2 \longrightarrow 1s^22s2p_{-1}$;
2. ${\rm Be^+}$: $1s^22s \longrightarrow 1s^22p_{-1}$;
3. Be: $1s^22s2p_{-1} \longrightarrow 1s^22p_{-1}3d_{-2}$.
4. ${\rm Be^+}$: $1s^22p_{-1} \longrightarrow 1s2p_{-1}3d_{-2}$;
5. Be: $1s^22p_{-1}3d_{-2} \longrightarrow 1s2p_{-1}3d_{-2}4f_{-3}$.
Crossovers (4) and (5) take place at relatively close values
of $\gamma$ and are not resolved in the figure.

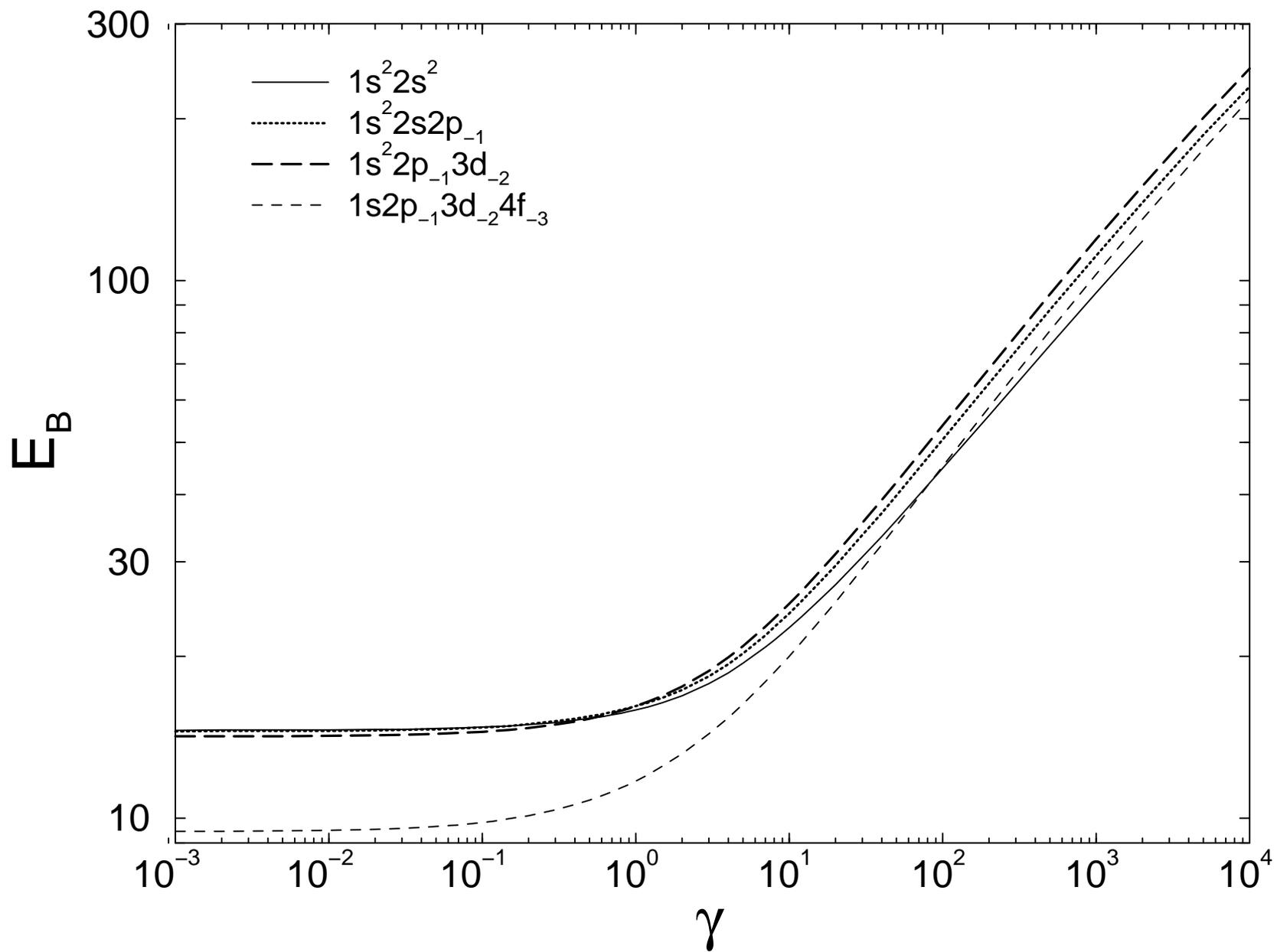
{\bf Figure 4.} The binding energies (in atomic units)
of the ground state electronic configurations
of the Be atom depending on the magnetic field strength.
The field strength is given in units of $\gamma=(\frac{B}{B_0})$,
$B_0=\hbar c/ea_0^2=2.3505 {\cdot} 10^5$T.

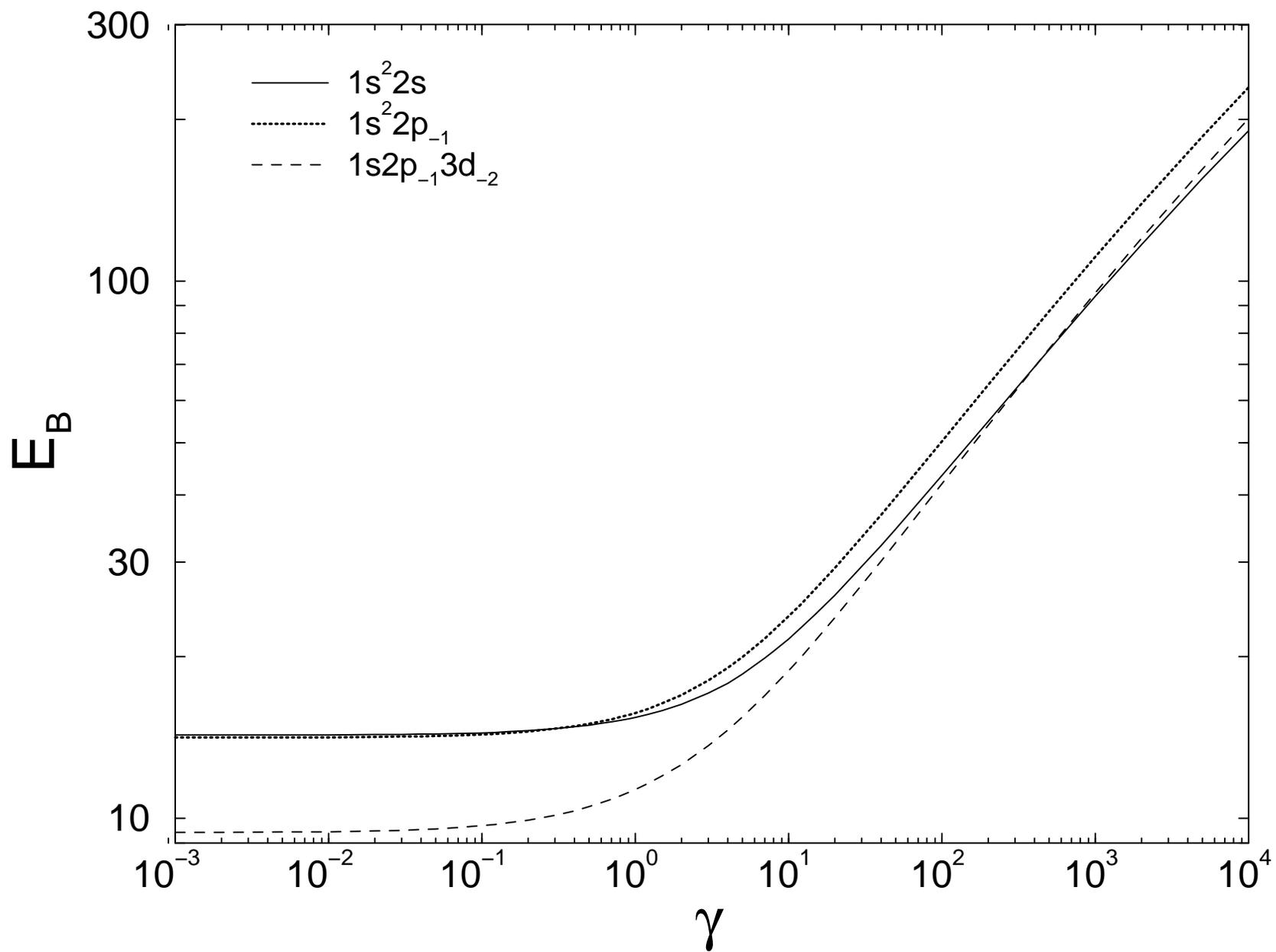
{\bf Figure 5.} The binding energies (in atomic units)
of the ground state electronic configurations
of the ${\rm Be^+}$ ion depending on the magnetic field strength.
The field strength is given in units of $\gamma=(\frac{B}{B_0})$,
$B_0=\hbar c/ea_0^2=2.3505 {\cdot} 10^5$T.

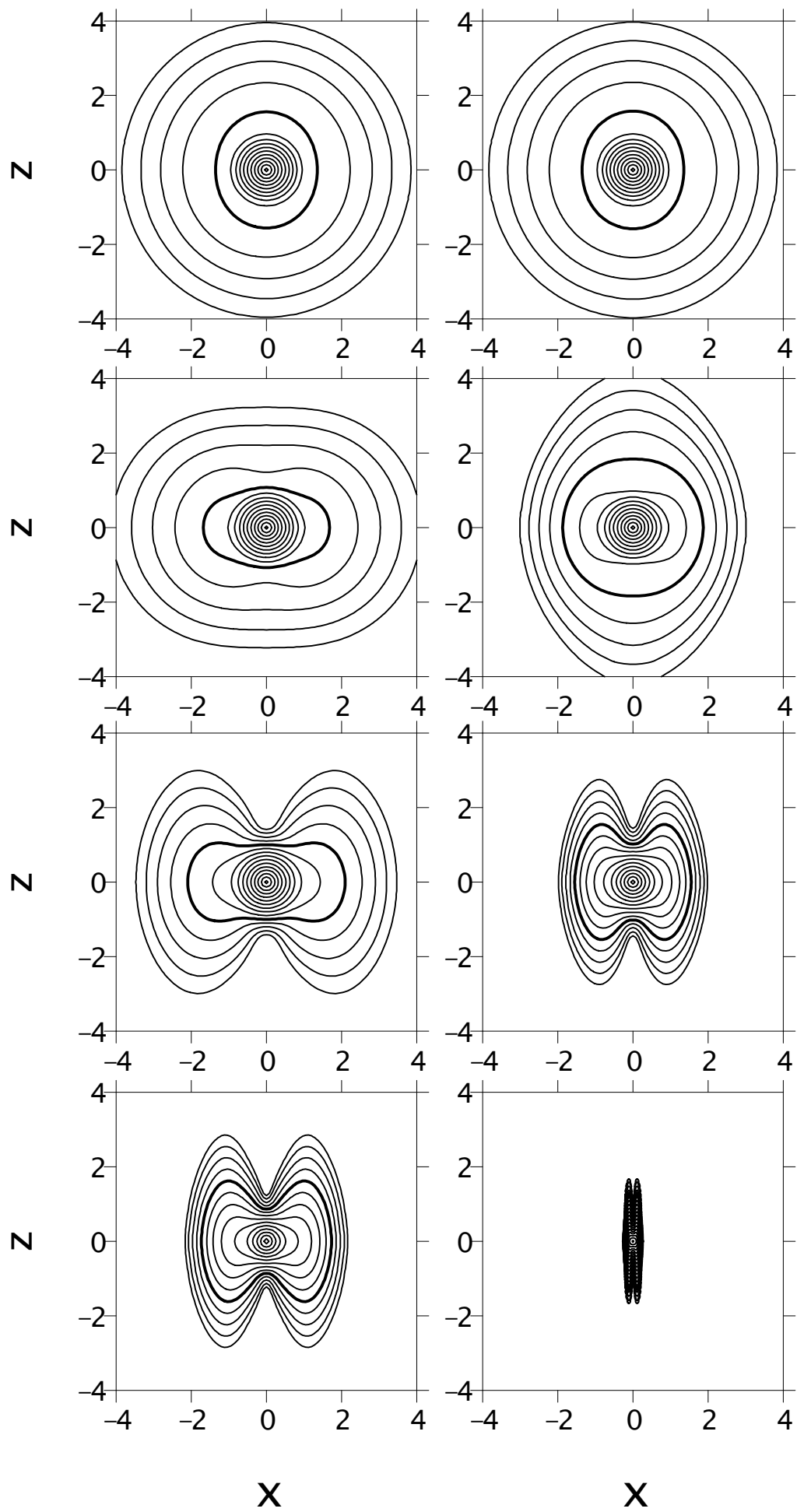
{\bf Figure 6.} Contour plots of the total electronic
densities for the ground state of the beryllium atom.
For neighbouring lines the densities are different by a factor of 2.
The coordinates $z$, $x$ as well as the
corresponding field strengths are given in atomic units.
Each row presents plots for a ground state configuration
at its lower (left) and upper (right) intersection points.
Rows: 1. $1s^22s^2$: $\gamma=0$ and $\gamma=0.0412$;
2. $1s^22s2p_{-1}$: $\gamma=0.0412$ and $\gamma=0.957$;
3. $1s^22p_{-1}3d_{-2}$: $\gamma=0.957$ and $\gamma=4.567$;
4. $1s2p_{-1}3d_{-2}4f_{-3}$: $\gamma=4.567$ and $\gamma=500$.

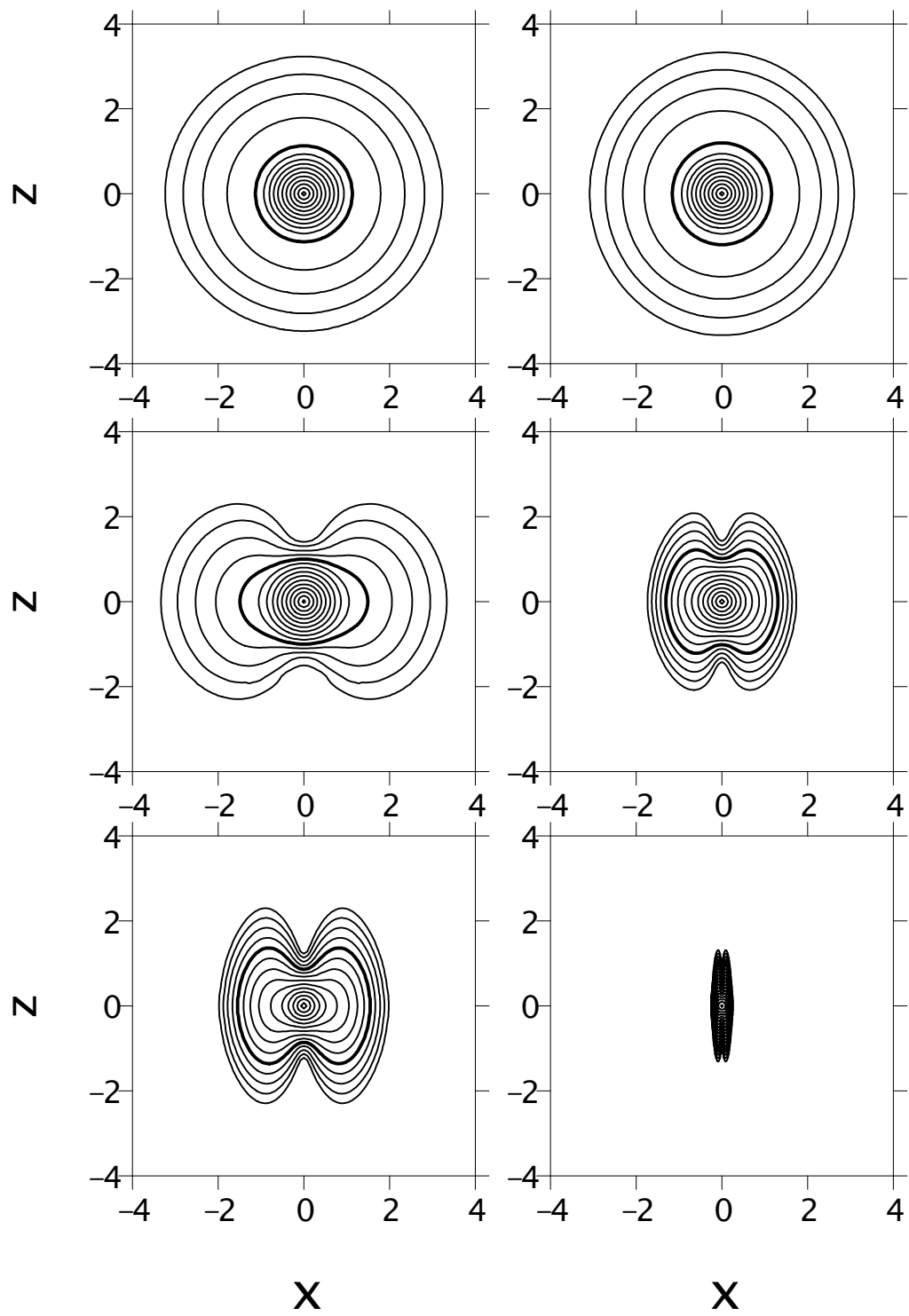
{\bf Figure 7.} Contour plots of the total electronic
densities for the ground state of the beryllium positive ion.
For neighbouring lines the densities are different by a factor of 2.
The coordinates $z$, $x$ as well as the
corresponding field strengths are given in atomic units.
Each row presents plots for a ground state configuration
at its lower (left) and upper (right) intersection points.
Rows: 1. $1s^22s$: $\gamma=0$ and $\gamma=0.3185$;
2. $1s^22p_{-1}$: $\gamma=0.3185$ and $\gamma=4.501$;
3. $1s2p_{-1}3d_{-2}$: $\gamma=4.501$ and $\gamma=500$.

\onecolumn

\newpage
\begin{table}
\caption{The Hartree-Fock ground state configurations of the beryllium atom
in external magnetic fields.
The configurations presented in the table are the ground state configurations
for $\gamma_{\rm min}\leq\gamma\leq\gamma_{\rm max}$. 
Atomic units are used.}
\begin{tabular}{@{}llllrllllll}
no.&$\gamma_{\rm min}$&$\gamma_{\rm max}$& The ground state configuration&$M$&$S_z$&$E(\gamma_{\rm min})$\\
\noalign{\hrule}
1&0     &0.0412   &$1s^2 2s^2 $             &$0 $&$0 $&$-14.57336$\\
2&0.0412&0.957    &$1s^22s2p_{-1}$          &$-1$&$-1$&$-14.57098$\\
3&0.957 &4.567    &$1s^22p_{-1}3d_{-2}$     &$-3$&$-1$&$-15.13756$\\
4&4.567 &$\infty$ &$1s2p_{-1}3d_{-2}4f_{-3}$&$-6$&$-2$&$-15.91660$\\
\end{tabular}
\label{tab:Begrcon}
\end{table}

\begin{table}
\caption{The Hartree-Fock ground state configurations of the ${\rm Be^+}$ ion
in external magnetic fields.
The configurations presented in the table are the ground state configurations
for $\gamma_{\rm min}\leq\gamma\leq\gamma_{\rm max}$.
Atomic units are used.}
\begin{tabular}{@{}llllrllllll}
no.&$\gamma_{\rm min}$&$\gamma_{\rm max}$& The ground state configuration&$M$&$S_z$&$E(\gamma_{\rm min})$\\
\noalign{\hrule}
1&0       &0.3185  &$1s^2 2s $        &$0 $&$-1/2$&$-14.27747$\\
2&0.3185  &4.501   &$1s^22p_{-1}$     &$-1$&$-1/2$&$-14.38602$\\
3&4.501   &$\infty$&$1s2p_{-1}3d_{-2}$&$-3$&$-3/2$&$-15.01775$\\
\end{tabular}
\label{tab:Bepgrcon}
\end{table}

\newpage
\begin{table}
\caption{The total energies of the ground state configurations 
of the beryllium atom
depending on the magnetic field strength.
The figures in parentheses are the labels of the ground state configurations
provided in the first column of table I. 
Atomic units are used.
}
\begin{tabular}{@{}llllllllll}
$\gamma$&$E(1)    $&$E(2)     $&$E(3)     $&$E(4)   $\\
\noalign{\hrule}
0.000 &$-14.57336 $&$-14.51206$&$-14.19023$&$-9.44321  $\\
0.001 &$-14.57336 $&$-14.51357$&$-14.1928 $&$-9.4483   $\\
0.002 &$-14.57335 $&$-14.51507$&$-14.1952 $&$-9.4532   $\\
0.005 &$-14.57332 $&$-14.51953$&$-14.2025 $&$-9.4675   $\\
0.01  &$-14.57322 $&$-14.52690$&$-14.2142 $&$-9.4903   $\\
0.02  &$-14.57279 $&$-14.54138$&$-14.2361 $&$-9.5331   $\\
0.03  &$-14.57209 $&$-14.55553$&$-14.2566 $&$-9.5735   $\\
0.04  &$-14.57111 $&$-14.56933$&$-14.27587$&$-9.6121   $\\
0.05  &$-14.56986 $&$-14.58281$&$-14.29437$&$-9.6493   $\\
0.07  &$-14.56657 $&$-14.60879$&$-14.32933$&$-9.7198   $\\
0.1   &$-14.55971 $&$-14.64548$&$-14.3780 $&$-9.82     $\\
0.12  &$-14.55395 $&$-14.66851$&$-14.40838$&$-9.8805   $\\
0.15  &$-14.54367 $&$-14.70108$&$-14.45145$&$-9.9692   $\\
0.2   &$-14.52261 $&$-14.75065$&$-14.51761$&$-10.1081  $\\
0.3   &$-14.46861 $&$-14.83520$&$-14.63369$&$-10.36220 $\\
0.3185&            &$-14.84905$\\
0.4   &$-14.40279 $&$-14.90464$&$-14.73396$&$-10.59384 $\\
0.5   &$-14.32860 $&$-14.96264$&$-14.82272$&$-10.80901 $\\
0.6   &$-14.24832 $&$-15.01171$&$-14.90262$&$-11.01121 $\\
0.7   &$-14.16352 $&$-15.05368$&$-14.97542$&$-11.20281 $\\
0.8   &$-14.07526 $&$-15.08989$&$-15.04232$&$-11.38545 $\\
0.9   &$-13.98431 $&$-15.12138$&$-15.10422$&$-11.56045 $\\
1.    &$-13.89120 $&$-15.14899$&$-15.16178$&$-11.72880 $\\
1.2   &$-13.69990 $&$-15.19498$&$-15.26583$&$-12.04863 $\\
1.5   &$-13.40329 $&$-15.24757$&$-15.39926$&$-12.49432 $\\
2.    &$-12.88908 $&$-15.30815$&$-15.57496$&$-13.16961 $\\
3.    &$-11.79811 $&$-15.36376$&$-15.79985$&$-14.35016 $\\
4.    &$-10.633617$&$-15.34275$&$-15.90161$&$-15.38050 $\\
4.501 &            &           &$-15.91626$             \\
5.    &$ -9.40602 $&$-15.25183$&$-15.91027$&$-16.30690 $\\
7.    &$ -6.79760 $&$-14.89530$&$-15.71644$&$-17.94005 $\\
8.    &$ -5.43095 $&$-14.64516$&$-15.53623$&$-18.67389 $\\
10.   &$ -2.5988  $&$-14.03046$&$-15.04644$&$-20.01753 $\\
12.   &$ +0.34064 $&$-13.29115$&$-14.41743$&$-21.23057 $\\
15.   &$ +4.9055  $&$-12.00063$&$-13.27286$&$-22.86513 $\\
20.   &$+12.8201  $&$ -9.49118$&$-10.97100$&$-25.23250 $\\
30.   &$+29.3964  $&$ -3.59324$&$ -5.40704$&$-29.11102 $\\
40.   &$+46.5935  $&$ +3.04026$&$ +0.95677$&$-32.28415 $\\
50.   &$+64.186   $&$+10.1472 $&$ +7.83395$&$-35.00768 $\\
100.  &$+155.286  $&$+49.4177 $&$+46.25962$&$-45.10519 $\\
200.  &$+343.899  $&$+135.659 $&$+131.4188$&$-58.08264 $\\
500.  &$+924.20   $&$+411.830 $&$+405.7027$&$-80.67357 $\\
1000. &$+1905.14  $&$+888.70  $&$+880.706 $&$-102.75480$\\
2000. &$+3881.5   $&$+1860.40 $&$+1850.052$&$-129.9790 $\\
5000. &$          $&$+4813.56 $&$+4799.35 $&$-175.2704 $\\
10000.&$          $&$+9770.37 $&$+9752.24 $&$-217.695  $\\
\end{tabular}
\label{tab:Beenergy}
\end{table}

\newpage
\begin{table}
\caption{The total energies of the ground state 
configurations of the ${\rm Be^+}$ ion
depending on the magnetic field strength.
The figures in parentheses are the labels of the ground state configurations
provided in the first column of table II. 
Atomic units are used.}
\begin{tabular}{@{}llllllllll}
$\gamma$&$E(1)   $&$E(2)      $&$E(3)    $\\
\noalign{\hrule}
0.000 &$-14.27747$&$-14.13093 $&$-9.41056$&$$\\
0.001 &$-14.27797$&$-14.13195 $&$-9.41358$&$$\\
0.002 &$-14.27846$&$-14.13294 $&$-9.41657$&$$\\
0.005 &$-14.27995$&$-14.13593 $&$-9.42551$&$$\\
0.01  &$-14.28241$&$-14.14087 $&$-9.44028$&$$\\
0.02  &$-14.28725$&$-14.15066 $&$-9.46939$&$$\\
0.03  &$-14.29198$&$-14.16030 $&$-9.49791$&$$\\
0.04  &$-14.29659$&$-14.16980 $&$-9.52587$&$$\\
0.0412&$-14.29714$                           \\
0.05  &$-14.30111$&$-14.17916 $&$-9.55332$&$$\\
0.07  &$-14.30981$&$-14.19746 $&$-9.60670$&$$\\
0.1   &$-14.32207$&$-14.22390 $&$-9.68356$&$$\\
0.12  &$-14.32972$&$-14.24088 $&$-9.73294$&$$\\
0.15  &$-14.34047$&$-14.26542 $&$-9.80463$&$$\\
0.2   &$-14.35648$&$-14.30406 $&$-9.91878$&$$\\
0.3   &$-14.38212$&$-14.37402 $&$-10.13144$&$$\\
0.4   &$-14.40046$&$-14.43599 $&$-10.32817$&$$\\
0.5   &$-14.41282$&$-14.49163 $&$-10.51259$\\
0.6   &$-14.42022$&$-14.54210 $&$-10.68705$\\
0.7   &$-14.42350$&$-14.58821 $&$-10.85323$\\
0.8   &$-14.42335$&$-14.63059 $&$-11.01236$\\
0.9   &$-14.42029$&$-14.66971 $&$-11.16540$\\
0.957 &           &$-14.69069 $            \\
1.    &$-14.41478$&$-14.70591 $&$-11.31312$\\
1.2   &$-14.39782$&$-14.77070 $&$-11.59490$\\
1.5   &$-14.36143$&$-14.85169 $&$-11.98978$\\
2.    &$-14.28225$&$-14.95181 $&$-12.59206$\\
3.    &$-14.08247$&$-15.05201 $&$-13.65352$\\
4.    &$-13.83797$&$-15.05004 $&$-14.58615$\\
4.567 &           &            &$-15.07310$\\
5.    &$-13.55019$&$-14.96820 $&$-15.42817$\\
7.    &$-12.85647$&$-14.61928 $&$-16.91814$\\
8.    &$-12.45821$&$-14.37080 $&$-17.58931$\\
10.   &$-11.57652$&$-13.75773 $&$-18.820184$\\
12.   &$-10.59993$&$-13.01900 $&$-19.93310$\\
15.   &$ -8.99386$&$-11.72840 $&$-21.43461$\\
20.   &$ -6.03364$&$-9.217910 $&$-23.612005$\\
30.   &$ +0.59244$&$-3.31723  $&$-27.18373$\\
40.   &$ +7.81557$&$+3.31895  $&$-30.10832$\\
50.   &$+15.4261 $&$+10.42836 $&$-32.61959$\\
100.  &$+56.5516 $&$+49.70820 $&$-41.93414$\\
200.  &$+145.1649$&$+135.95916$&$-53.90638$\\
500.  &$+425.471 $&$+412.14745$&$-74.73619$\\
1000. &$+906.37  $&$+889.0264 $&$-95.07513$\\
2000. &$+1883.08 $&$+1860.7100$&$-120.11947$\\
5000. &$+4844.6  $&$+4814.005 $&$-161.7052$\\
10000.&$+9809.3  $&$+9770.643 $&$-200.5709$\\
\end{tabular}
\label{tab:Bepenergy}
\end{table}

\end{document}